\documentclass[reprint, groupedaddress,aps
]{revtex4-2}

\usepackage{hyperref}
\usepackage{blkarray}
\usepackage{soul}
\usepackage{comment}
\usepackage{graphicx}% Include figure files
\usepackage{dcolumn}% Align table columns on 
\usepackage{booktabs} % For formal tables
\usepackage{amsmath,amssymb}
\usepackage{color}

\hypersetup{
    colorlinks = true,
    linkcolor = blue,
    citecolor = blue,
    filecolor = blue,
    urlcolor = blue
}

\DeclareMathOperator{\csch}{csch}

\makeatother
\begin{document}

\preprint{APS/123-QED}

\title{Heisenberg machines with programmable spin-circuits}% Force line breaks 
\author{Saleh Bunaiyan$^{1,3}$, Supriyo Datta$^2$, and Kerem Y. Camsari$^3$}

\affiliation{$^1$Electrical Engineering Department, \\ King Fahd University of Petroleum $\&$ Minerals (KFUPM), Dhahran 31261, Saudi Arabia}

\affiliation{$^2$Elmore Family School of Electrical and Computer Engineering, Purdue University, West Lafayette, Indiana 47907, USA}%

\affiliation{$^3$Department of Electrical and Computer Engineering,\\ University of California at Santa Barbara, Santa Barbara, CA 93106, USA}

\date{\today}

\begin{abstract}
We show that we can harness two recent experimental developments to build a compact hardware emulator for the classical Heisenberg model in statistical physics. The first is the demonstration of spin-diffusion lengths in excess of microns in graphene even at room temperature. The second is the demonstration of low barrier magnets (LBMs) whose magnetization can fluctuate rapidly even at sub-nanosecond rates. Using experimentally benchmarked circuit models, we show that an array of LBMs driven by an external current source has a steady-state distribution corresponding to a classical system with an energy function of the form $E = -1/2\sum_{i,j}  J_{ij} (\hat{m}_i \cdot \hat{m}_j$). This may seem surprising for a non-equilibrium system but we show that it can be justified by a Lyapunov function corresponding to a system of coupled Landau-Lifshitz-Gilbert (LLG) equations. The Lyapunov function we construct describes LBMs interacting through the spin currents they inject into the spin neutral substrate. We suggest ways to tune the coupling coefficients $J_{ij}$ so that it can be used as a hardware solver for optimization problems involving continuous variables represented by vector magnetizations, similar to the role of the Ising model in solving optimization problems with binary variables. Finally, we train a Heisenberg XOR gate based on a network of four coupled stochastic LLG equations, illustrating the concept of probabilistic computing with a programmable Heisenberg model. 
\end{abstract}

\maketitle

\section{Introduction}
\label{Introduction}
With the slowing down of Moore's law, there is tremendous interest in unconventional computing approaches. Prominent among these approaches are the energy-based models (EBMs) inspired by statistical physics, in which the problem is mapped to an energy function, such that the solution corresponds to the minimum energy state \cite{Hinton_EBM,Berloff_XY_realization,Masoud_EBM}, which can then be identified using powerful optimization algorithms.

A particularly attractive way to find the minimum energy state of an energy function becomes possible if we can design a physical system whose natural physics makes it relax to this state. Such a system could be harnessed to solve this class of problems orders of magnitude faster and more efficiently than an algorithm implemented using conventional transistors. The main reason behind this gap is the extensive costs of designing stochastic spins in deterministic hardware where it  takes tens of thousands of transistors to emulate the necessary tunable randomness \cite{kobayashi2023cmos+}. 
An interesting EBM is the classical Heisenberg model
\begin{equation}
   E = - \frac{1}{2} \sum_{i,j} J_{ij}\,(\hat{m}_i\cdot\hat{m}_j)
   \label{Heisen}
\end{equation}

\noindent where $\hat{m}_{i,j}$ are unit vectors in 3D, which is different from the Ising model
\begin{equation}
   E_{\mbox{{\footnotesize ising}}} = - \frac{1}{2} \sum_{i,j}  J_{ij} \ s_i s_j
   \label{Ising}
\end{equation}

 \noindent where $s_{i}, s_{j}$ are variables with only two allowed values, $\pm 1$. Hopfield networks and Boltzmann machines based on the Ising model, Eq. \eqref{Ising}, have generated tremendous recent excitement and even mapped to physical systems \cite{Borders2019_experimental_pbit,mohseni2022ising}. The modern Hopfield network \cite{ramsauer2021hopfield,krotov2023new,Widrich2020Modern}, based on the classical Heisenberg model, Eq. \eqref{Heisen}, shows promise in machine learning applications and it should be of great interest to map it to an energy-efficient physical system.

\begin{figure}[t!]
   \centering
    \includegraphics[width=\linewidth]{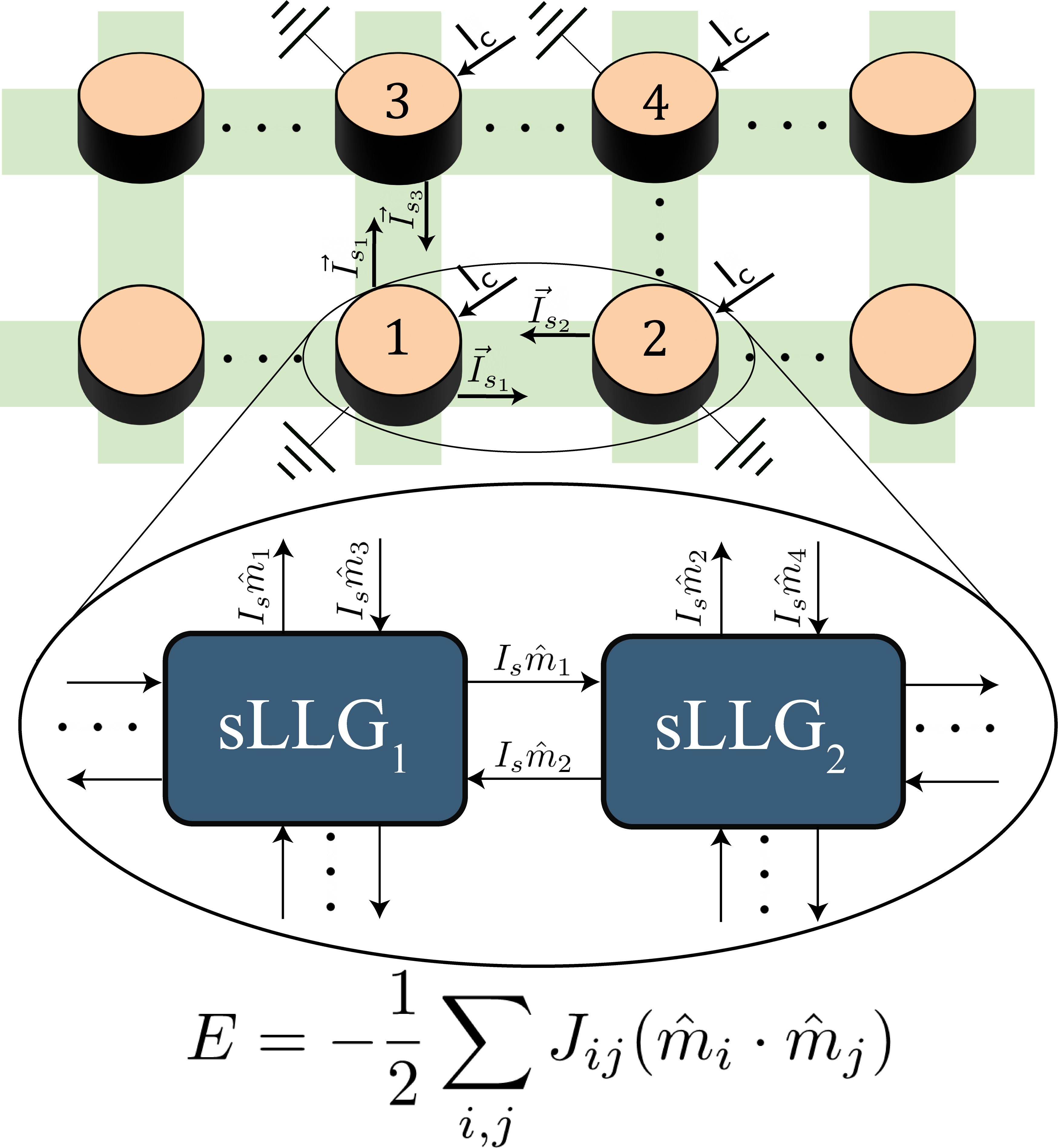}
    \caption{\textbf{Heisenberg machines with programmable spin-circuits.} Low barrier magnets (LBMs) and the spin neutral channels form the Heisenberg array, where an external charge current is injected into the LBMs and thereby inducing a spin current inside the channel. Magnets are assumed to have zero energy barrier with no effective anisotropy. The proposed array is analogous to a system of coupled stochastic Landau–Lifshitz–Gilbert (sLLG) equations where each sLLG  sends a spin current vector $\vec{I}_{s}$ to its neighbors, where the spin current is defined as a function of the magnet magnetization $\vec{I}_{si} = I_s \hat m_i$ . This coupled LBM system minimizes the Heisenberg Hamiltonian with continuous spin states, functioning as a Heisenberg machine, analogous to Ising machines.}
    \label{figure_1}
\end{figure}

In this work, we propose and establish the feasibility of mapping Eq.~\eqref{Heisen} to a physical system consisting of an array of low barrier magnets (LBMs) interacting via spin currents through a spin-neutral channel (e.g., Cu, graphene) as shown in Fig.~\ref{figure_1}. The system is similar to the 2D graphene channels with long spin-diffusion length ($\lambda_s$) used to demonstrate room temperature spin logic \cite{ishihara2020gate,panda2020ultimate,bisswanger2022cvd,Saroj_Dash_spin_experiment}, but with one $\textit{key difference}$. Here, the magnets are not the usual stable magnets, but LBMs similar to those used to represent Ising spins or probabilistic bits (p-bits) \cite{camsari2017stochastic,camsari2017Implementing}, that have  been shown to fluctuate with GHz rates \cite{Nanosecond_Fukami,Nanosecond_Kaiser,Schnitzspan2023Nanosecond}. A charge current is driven through each LBM $i$ by an external source, and the associated spin current diffuses through the substrate and exerts a spin-torque on a neighboring LBM $j$, leading to an effective interaction term $J_{ij}$ between them.

Note that this is a particularly compact physical realization of Eq. \eqref{Heisen} compared to the reported realizations of the Ising model (Eq. \eqref{Ising}) using p-bits \cite{Borders2019_experimental_pbit,Kaiser_experimental_pbit,Andrea_exp_MTJ}, which require separate hardware to implement ``synapses" that sense the state of each p-bit and drive neighboring p-bits accordingly. The term synapses here is used to refer to the coupling strength (weights) between neurons, where we assume that synapses have fixed values. Hence, the structure proposed here (Fig.~\ref{figure_1}) integrates both the neuron and the synapse at once, thus making it possible to implement relatively fast synapses that can utilize the nanosecond fluctuations that have been demonstrated \cite{Nanosecond_Fukami,Nanosecond_Kaiser,Schnitzspan2023Nanosecond} and could be designed to be even faster. The only additional hardware needed here is a final readout of the output magnets, rather than repeated readouts and re-injections from intermediate magnets.

The $\textit{central result}$ of this paper is to establish that the structure in Fig.~\ref{figure_1} indeed minimizes a classical Heisenberg energy function of the form Eq.~\eqref{Heisen}. This is not at all obvious since our structure is \emph{not} at equilibrium and is not expected to necessarily obey a Boltzmann law $\propto \exp(-E/k_BT)$, $k_B T$, being the thermal energy. What we theoretically show is that an energy function of the form Eq.~\eqref{Heisen} constitutes a Lyapunov function that is minimized by the LBM dynamical equations. Our numerical simulations further indicate that different configurations also follow a Boltzmann law at least approximately.

\section{Lyapunov functions}
\label{Lyapunov functions}
To model a structure like Fig.~\ref{figure_1} with $N$ LBMs we start from a set of $N$ coupled Landau-Lifshitz-Gilbert (LLG) equations using experimentally benchmarked physical parameters from \cite{Camsari_modular_approch,Sayed2016_spin_circuit_model,Saroj_Dash_spin_experiment}, which were carefully derived from the work of Bauer, Brataas and Kelly on magnetoelectric circuit theory \cite{Bauer}.  These were later turned into explicit circuit models that are modularly simulated in standard circuit simulators \cite{srinivasan2013modeling,manipatruni2012modeling,Camsari_modular_approch}. Throughout this work, we focus on LBMs with no shape or easy-plane anisotropy which can be practically built by reducing the energy barriers of magnets with perpendicular anisotropy (PMA). Similar to other dissipative dynamical systems \cite{graham1984existence}, we seek to establish a pseudo-energy function that is minimized over the time evolution of the system. We first show that a single magnet ($\hat{m}_1$) with an injected spin current $\vec{I}_s$, coming from another fixed magnet $\hat m_2$ such that $\vec{I}_s = I_s \hat{m}_2$, minimizes the function
\begin{equation}
   E = - \left({\frac{{I_s}}{I_0}}\right) \hat{m_1} \cdot \hat m_2 = - J_{12} \ \hat m_1 \cdot \hat m_2
   \label{eq: LBM_energy}
\end{equation}
\noindent where $J_{12}$ is the dimensionless and symmetric coupling ($J_{12}=J_{21}$) coefficient defined as $I_s/I_0$. $I_0$ is a constant given by $2q \alpha k_B T /\hbar$ where $q$ is the electron charge, $\hbar$ is the reduced Planck's constant and $\alpha$ is the damping coefficient of the nanomagnet. Next, we first show how $I_0$ can be formally derived from the Fokker-Planck equation (FPE).

\subsection{Fokker-Planck Equation for Coupled LBMs}
\label{fpe}

Suppose that the spins current  $\vec{I}_s$ injected into $\hat m_1$ is polarized in the direction of another fixed magnet, $\hat m_2$,
such that $\vec{I}_s$ = ${I}_s \hat m_2$. We will assume $\hat m_2 = \hat z$ without loss of generality; any arbitrary direction would produce the same results, but for the sake of simplifying the math, we chose that direction. For this system, the following  FPE can be written at steady-state \cite{brown1963thermal}:

\begin{equation}
  \left(\frac{\alpha \gamma k_B T}{(1+\alpha^2) M_s   \mathrm{Vol.}}\right) \displaystyle \nabla^2 P(\theta_1,\phi_1) - \nabla \cdot \left(P(\theta_1,\phi_1)\frac{d\hat{m}_1}{dt}\right) = 0
  \label{eq: fpe}
\end{equation}
where $\gamma$ is the gyromagnetic ratio of the electron, $M_s$ is the saturation magnetization of the magnet, $\mathrm{Vol.}$ is the volume of the magnetic body. The FPE derivation discussed in the seminal work of Brown \cite{brown1963thermal} naturally does not consider the effect of the spin-transfer-torque since this phenomenon was discovered later. Here, we use a modified FPE that expands Brown's work to include the spin-transfer-torque effect, following the approach discussed in \cite{li2004thermally}.

At steady-state, the magnetization $\hat m_1$ follows a Boltzmann-like law according to the energy, $E$, defined by Eq.~\eqref{eq: LBM_energy}:  

\begin{equation}
    P(\theta_1,\phi_1) = \frac{1}{Z} \exp ( - E ) 
    \label{eq: FPE solution}
\end{equation}

In turn, $d \hat m_1/dt$ can be obtained from the deterministic LLG equation \cite{Butler2012Switching} for $\hat{m}_1$:
\begin{equation}
   (1+\alpha^2)\frac{d\hat{m}_1}{dt} = \frac{\hat{m}_1\times(\vec{I_s}\times\hat{m}_1)}{qN_s} + \frac{\alpha(\hat{m}_1\times\vec{I_s})}{qN_s}
   \label{eq: LLG}
\end{equation}
where $N_s$ is the number of spins in the magnetic volume, i.e., $N_s = M_s \mathrm{Vol.}/\mu_B$, $\mu_B$ being the Bohr magneton. Direct substitution of Eq. \eqref{eq: FPE solution} into Eq. \eqref{eq: fpe} shows, after several steps of tedious algebra, the solution satisfies the Fokker-Planck equation. Next, we show that $dE/dt$ is always negative for coupled LBMs.

\subsection{Lyapunov Functions: Two-Magnet System}

Armed with our result in Eq.~\eqref{eq: LBM_energy}, we consider Lyapunov function for two coupled LBMs ($N=2$). The main idea is to find an ``energy'' whose rate of change is nonpositive:
\begin{equation}
    \frac{dE}{dt} = \vec{\nabla}_{\hat m}E \cdot \frac{d\hat{m}}{dt} \leq 0
    \label{Lyapunov_condition}
\end{equation}

Consider the same scenario we started from in Eq.~\eqref{eq: fpe}: magnet $\hat m_1$ is receiving a spin current, $\vec{I}_s$, polarized in the direction of another fixed magnet, $\hat m_2$, such that $\vec{I}_s=I_s \hat m_2 = (I_0)J_{12} \hat m_2$. $J_{12}$ is the dimensionless coupling between magnet 2 and magnet 1. For this system, define the Lyapunov function, $E$, as: 

\begin{equation}
E = -\left(\frac{I_s}{I_0}\right) \hat m_1 \cdot \hat m_2 \quad \mbox{with} \quad  \vec{\nabla}_{\hat m_1} E = -\left(\frac{I_s}{I_0}\right)\hat m_2 = -\frac{\vec{I_s}}{I_0}
\label{eq:Lyapunov_2}
\end{equation}
From the LLG equation for $\hat m_1$  (Eq.~\eqref{eq: LLG}), we obtain: 
\begin{equation}
   C\frac{d\hat{m}_1}{dt} = \vec{I_s} - \hat{m}_1(\hat{m}_1\cdot \vec{I_s})+\alpha(\hat{m}_1\times \vec{I_s})
   \label{LLG_single_magnet_constant_C}
\end{equation}
where we defined $ C = q N_s(1+\alpha^2) > 0 $. Combinations of Eq. \eqref{LLG_single_magnet_constant_C} and Eq. \eqref{eq:Lyapunov_2} using Eq.~\eqref{Lyapunov_condition} yields:

\begin{equation}
   C\frac{dE}{dt} = -\frac{\vec{I_s}}{I_0} \cdot [\vec{I_s} - \hat{m}_1(\hat{m}_1\cdot \vec{I_s})+\alpha(\hat{m}_1\times \vec{I_s})] 
   \label{LLG_single_magnet_Lyapunov}
\end{equation}

Finally by simplifying Eq. \eqref{LLG_single_magnet_Lyapunov} we can show that the condition for energy minimization is satisfied:

\begin{equation}
   C\frac{dE}{dt} = -\frac{[\vec{I_s} \cdot \vec{I_s} - (\hat{m}_1\cdot \vec{I_s})^2]}{I_0} \leq 0
   \label{single_magnet_Lyapunov_condition}
\end{equation}

\subsection{Lyapunov Function: $N$-Magnet System}

Using the same approach, we extend this result to a system of $N$ coupled LBMs, described by $N$ coupled LLG equations: 
\begin{equation}
   (1+\alpha^2)\frac{d\hat{m}_i}{dt} = \frac{\hat{m}_i\times( I_0\,\vec{I_i}\times\hat{m}_i)}{qN_s} + \frac{\alpha(\hat{m}_i\times I_0\,\vec{I_i})}{qN_s}
   \label{eq: N-LLG}
\end{equation}
with the input to the $i^{\rm th}$ magnet defined as $\vec{I_i} = \sum_{j}J_{ij}\hat{m}_j$. Given how the Lyapunov function for the 2-magnet system can be formally derived to follow a Boltzmann-like equation, we posit the following Lyapunov function for the $N$-magnet system: 

\begin{equation}
   E = - \frac{1}{2}\sum_{i,j} J_{ij}\,(\hat{m}_i\cdot\hat{m}_j) = - \frac{1}{2}\sum_{i,j} \,\frac{I_{sij}}{I_0}\,(\hat{m}_i\cdot\hat{m}_j)
   \label{Lyapunov_N_magnet}
\end{equation}
Assuming the reciprocity of the interaction strengths ($J_{ij} = J_{ji}$):
\begin{equation}
  \vec{\nabla}_{\hat m_i}E = - \sum_{j}J_{ij}\hat{m}_j = -\vec{I_i}
   \label{gradient_N_magnet}
\end{equation}

The coupled LLG equations for $N$-magnets can be written as:

\begin{equation}
   C\frac{d\hat{m}_i}{dt} =I_0[\vec{I_i} - \hat{m}_i(\hat{m}_i\cdot \vec{I_i})+\alpha(\hat{m}_i\times \vec{I_i})]
   \label{LLG_N_magnet_constant_C}
\end{equation}
Combinations of Eq. \eqref{LLG_N_magnet_constant_C} and Eq. \eqref{gradient_N_magnet} with Eq. \eqref{Lyapunov_condition} yields:

\begin{equation}
    C\frac{dE}{dt} = \sum_{i}-(\vec{I_i} \cdot I_0[\vec{I_i} - \hat{m}_i(\hat{m}_i\cdot \vec{I_i})+\alpha(\hat{m}_i\times \vec{I_i})])
    \label{LLG_N_magnet_Lyapunov}
\end{equation}
by noting that ($\sum_{i} \vec{I_i} \cdot \alpha(\hat{m}_i\times \vec{I_i}) = 0$), the expression can be further simplified to: 

\begin{equation}
   C\frac{dE}{dt} = \sum_{i}-I_0[\vec{I_i} \cdot \vec{I_i} - (\hat{m}_i\cdot \vec{I_i})^2] \leq 0
\end{equation}
which shows that the condition for energy minimization is also satisfied for $N$-magnets.

\begin{figure}[t!]
   \centering
    \includegraphics[width=\linewidth]{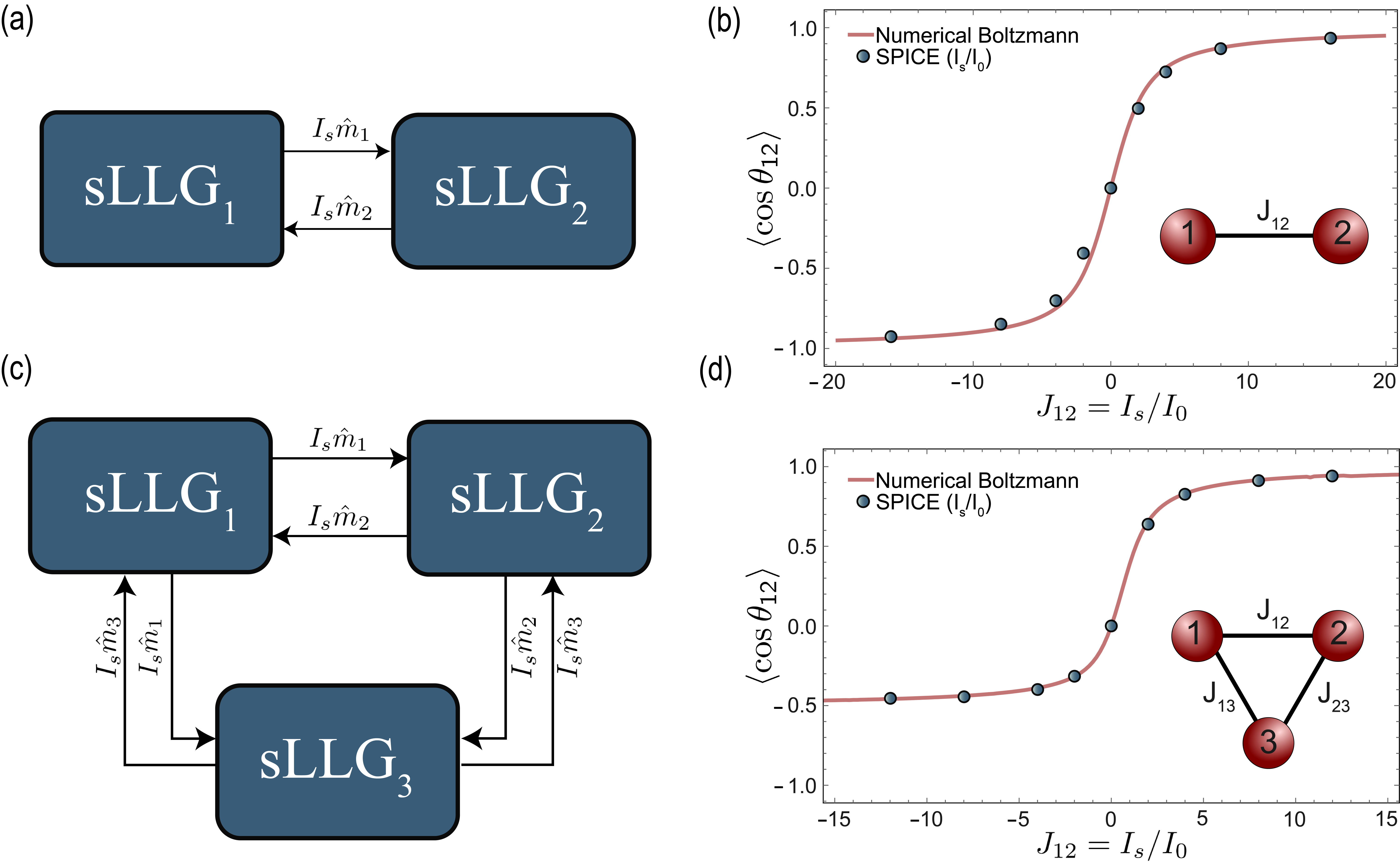}
    \caption{\textbf{Coupled LBMs with pure spin currents.} (a) Two coupled sLLG equations, where each sLLG sends a spin current along its magnetization direction $\vec{I}_{si}=I_s\hat{m}_i$ to the other sLLG. (b) The two coupled sLLG equations are contacted to the Boltzmann numerical solution by the predefined constant $I_0$. (c) Three coupled sLLG equations in a frustrated configuration. (d) The three coupled sLLG equations also match with the Boltzmann law, note that $I_0$ is configuration independent.}
    \label{figure_2}
\end{figure}

The above discussions imply $dE/dt \leq 0$, where $E$ is now the classical Heisenberg model described in Eq.~\eqref{Heisen}. As a result, in the \textit{deterministic limit}, the system dynamics tend to minimize the energy of a classical Heisenberg model whose parameters $J_{ij}$ can be programmed. As we will show later, in the presence of uncorrelated white noise (see stochastic LLG equation in Appendix \ref{Spin-circuit modules}) and transitions between all states with finite probability (i.e.,  ergodicity), the tendency to minimize energy leads the system to sample from the Boltzmann distribution with occasional energy-increasing state transitions enabled by thermal noise. This description of a non-equilibrium system with an equilibrium model is reminiscent of spin-currents interacting with PMA magnets \cite{Butler2012Switching}. 

\section{Numerical simulation}
\label{Numerical simulation}

In this section will present numerical results for two examples: a system of two LBMs and a frustrated system of three LBMs, indicating that the results follow the Boltzmann law, where states are sampled according to $\rho(\hat m_1, \hat m_2, \ldots \hat m_N)\propto \exp[-E(\hat m_1, \hat m_2, \ldots \hat m_N)]$.

\begin{figure*}[t!]
   \centering
    \includegraphics[width=7.15in]{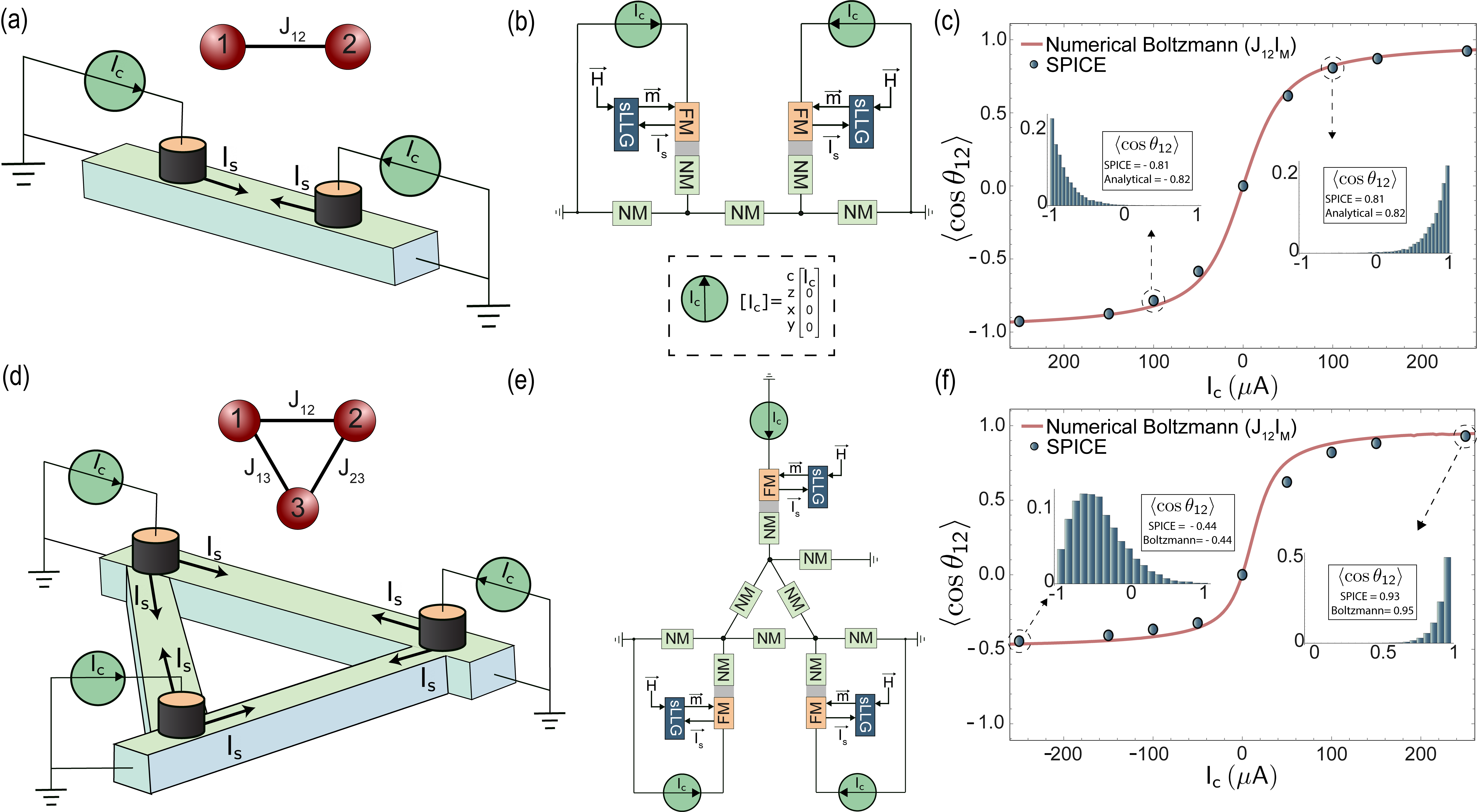}
    \caption{\textbf{Simulated hardware for the Heisenberg emulator}. (a) 2-LBM system that emulates a Heisenberg model of two spins. (b) Simulated configuration in HSPICE using benchmarked circuit models. Transport physics from the LBM to the channel and from the channel to the ground are characterized by an FM$|$NM and NM modules, respectively. Magnetization dynamics of the LBM are described by the sLLG modules. All circuits are described by 4 components: charge, and spins $(z, x, y)$. (c) The Boltzmann solution is related to the proposed hardware through $I_M$ that takes the charge-to-spin current conversions into account. The relation between spin current and interaction strength is given by ($J_{12} = I_s/I_0$). (d) 3-LBM system that emulates a Heisenberg model of three spins in a frustrated configuration. (e) Simulated configuration in HSPICE. (f) Boltzmann solution for the 3-magnet system. We use the same $I_M$ value in the 2-magnet system (even though a new charge-to-spin mapping may need to be measured or calculated, see Appendix \ref{Mapping_Charge_to_Spin}).  Insets show the probability distribution of the magnets correlation $\cos\theta_{12}$ at a given input $I_c$, which can be analytically derived for two LBMs by the aid of Eq. \eqref{Langevin_equation}. 
    }
    \label{figure_3}
\end{figure*}

Simulations were carried out on a standard circuit simulator (HSPICE) using 4-component spin-circuits \cite{Camsari_modular_approch,torunbalci2018modular}. The full circuits used for simulating the two examples are shown in Figs. \ref{figure_2} and \ref{figure_3}. The used modules are of two categories: transport and magnetism, where the transport through LBMs into the spin-neutral channel are characterized by an FM$|$NM interface module, which defines the interface between the ferromagnet (FM) and the normal metal (NM) \cite{Camsari_modular_approch,Bauer}. The magnetic fluctuations of LBMs are described by the stochastic LLG (sLLG) module, carefully benchmarked against corresponding FPEs for monodomain magnets \cite{torunbalci2018modular,Butler2012Switching}. The spin-neutral channel between LBMs is solely described by the transport module NM. All transport modules are represented by 4$\times$4 matrices describing the interactions between charge and spins in the $z,x,y$ directions. Further details on the spin-circuit modules, sLLG and thermal noise are discussed in Appendix \ref{Spin-circuit modules}. The physical parameters we used are reported in Appendix \ref{Physical Parameters}.

To show that a system of coupled LBMs sample from the classical Heisenberg model, we start by pure sLLG network that resembles the system of coupled LBMs. All numerical examples are based on the sLLG equation with the corresponding FPE given by Eq. \eqref{eq: fpe}. 

\subsection{Coupled LBMs without Transport}
\label{Coupled LBMs without Transport}

In Fig. \ref{figure_2}, we numerically study a system of two coupled sLLG equations, and we compare its response with numerical solution of Heisenberg Hamiltonian based on the Boltzmann law. We study the system by observing the correlation between the two coupled LBMs, which can be described by the cosine of the relative angle between their magnetization vectors $\langle \cos\theta_{12} \rangle$. In the two configurations of Fig. \ref{figure_2}, we compute correlations between the magnets purely based on spin-currents, for the full proposed hardware more complex analysis is needed.

If the two LBMs are coupled by a $J_{ij}$, their correlation can be computed from a Boltzmann-like equation using Eq.~\eqref{Heisen}:   
\begin{align}
\langle \cos\theta_{12} \rangle = \int_{S_1}\int_{S_2} \frac{1}{Z} \cos\theta_{12} \exp(-E) dS_1 dS_2
\label{2Magnets}
\end{align}
where ($dS_i = \sin(\theta_i)d\theta_i d\phi_i$) describes integration on the surface of the unit sphere and $Z$ is the partition function. For any given symmetric coupling $J_{ij}$ between LBM 1 and 2, we can evaluate Eq.~\eqref{2Magnets} numerically. Note that for 2-magnet system the average correlation can be also obtained through solving for the average magnetization $\langle m_z \rangle$ for a single magnet: 

\begin{equation}
\langle m_z \rangle = \frac{\displaystyle\int_{\theta=0}^{\theta=\pi}\int_{\phi=0}^{\phi=2\pi} \cos(\theta)/Z \exp\left(\frac{Is}{I_0}\cos(\theta) \right) \sin(\theta) \,d\theta d\phi}{\displaystyle\int_{\theta=0}^{\theta=\pi}\displaystyle\int_{\phi=0}^{\phi=2\pi}\displaystyle 1/Z \exp\left(\frac{Is}{I_0}\cos(\theta) \right) \sin(\theta) \,d\theta d\phi} 
\end{equation}
which is found to be the Langevin function:

\begin{equation}
    \langle m_z \rangle = \langle \cos\theta \rangle = \coth\left(\displaystyle\frac{I_s}{I_0}\right) - \displaystyle\frac{I_0}{I_s}
    \label{Langevin_equation}
\end{equation}
The Langevin function exactly overlaps with the Boltzmann numerical solution of $\langle \cos\theta_{12} \rangle$ shown in Fig.~\ref{figure_2}(b). This can be viewed as measuring the relative angle between the two magnets when one of the magnets is fixed to $+\hat z$ and the other magnet is free to move. This trick works due to the spherical symmetry of the system, discussed in more detail in Appendix \ref{Mapping_Charge_to_Spin}. 

We then study a frustrated system of three magnets, with the network configuration shown in Figs.~\ref{figure_2}(c) and ~\ref{figure_2}(d), described by the Boltzmann law as: 
\begin{equation}
\langle \cos\theta_{12} \rangle =  \int_{S_1}\int_{S_2}\int_{S_3} \frac{1}{Z} \cos\theta_{12} \exp(-E) dS_1 dS_2 dS_3
\label{eq: 3LBM-correlation}
\end{equation}
Our numerical results in Fig. \ref{figure_2} indicate that the two and three magnet systems are well-described by
the Boltzmann law, indicating that the network is actually sampling from the classical Heisenberg model. Moreover, it is important to note the generality of the constant $I_0$, which could guide the experimental realization of these network.

\subsection{Programmable Spin-Circuits}
The proposed physical structure of the two studied systems are described in Figs. \ref{figure_3}(a) and \ref{figure_3}(d). All magnets were chosen to be identical and receive the same input current $I_c$. In our setup, the charge current $I_c$ flows to a nearby ground injecting pure spin-currents in all directions, as commonly done in non-local spin valves \cite{kimura2006switching,Saroj_Dash_spin_experiment}.

To make contact between the circuit shown in Fig.~\ref{figure_3}(a) and Eq.~\eqref{2Magnets} we need the same normalizing parameter $I_0 = I_{sij}/J_{ij}$ where $I_{sij}$ is the component of the spin-current along magnet $j$ incident to magnet $i$. In addition to that, we need another configuration-dependent parameter relating the injected charge current to the spin-currents. We define a single parameter, $I_M$ that combines the two such that $J_{ij}= (I_c/I_M)$. For the 2-magnet system, we can find $I_{sij}$ analytically by defining a new basis: 
 ($\hat m_1,\hat m_2, \hat m_1 \times \hat m_2$) similar to the approach used in \cite{datta2011voltage}. By a clever coordinate transformation, we exactly solve for $I_M$ in the 2-magnet system in Appendix \ref{Mapping_Charge_to_Spin}. Numerical results from our spin-circuits match those obtained from our analytical solution, as shown in Fig.~\ref{figure_3}, where the analytical solution was found by substituting $J_{12} = I_c/I_M$ in Eq. \eqref{Langevin_equation}. Note that in general systems ($N>2$), $I_M$ may depend on the geometry of the channel and it may need to be found by experimental calibration in different systems. 

\begin{figure}[t!]
   \centering
    \includegraphics[width=3.25in]{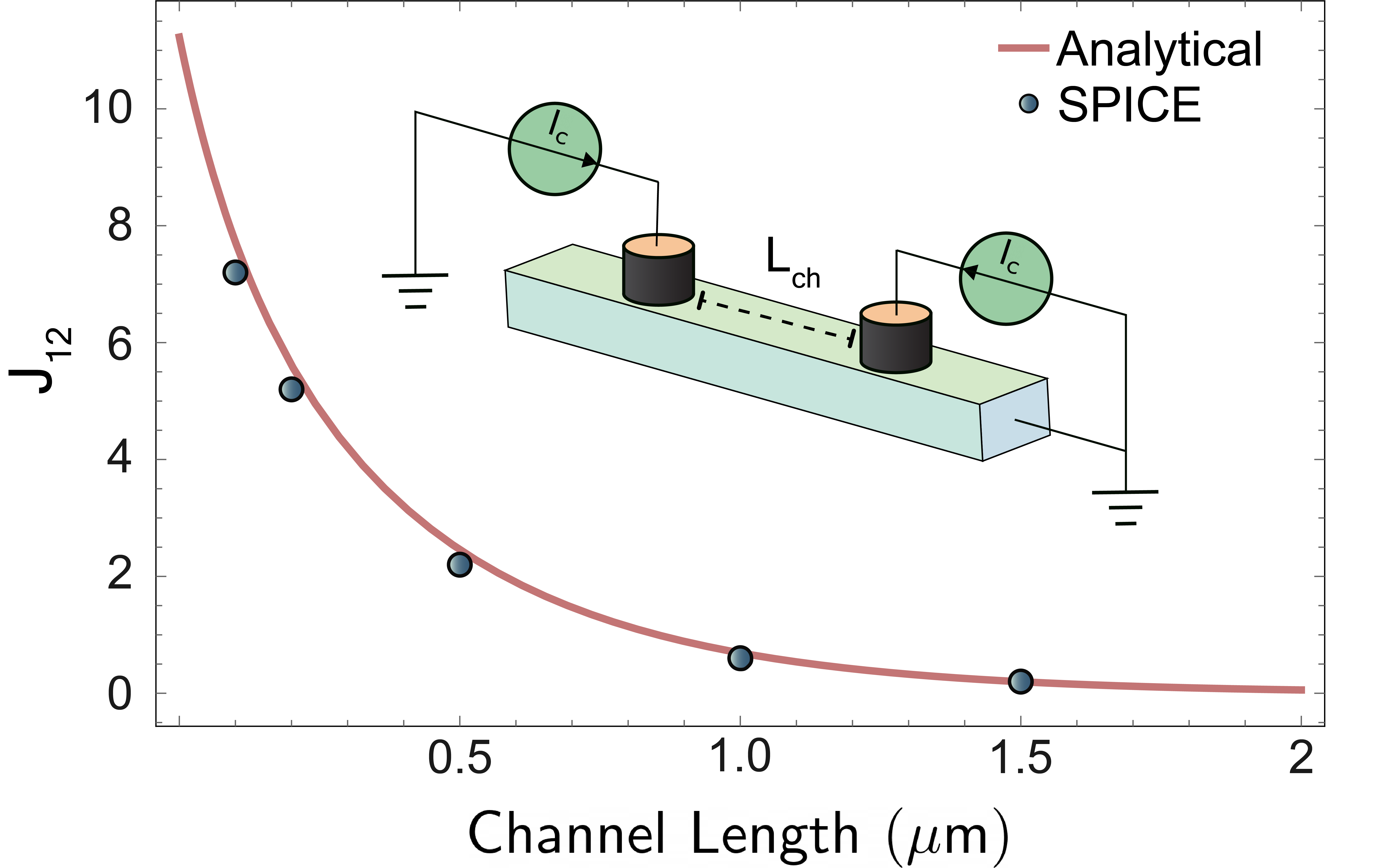}
    \caption{\textbf{Programmability of the interaction strength $J_{ij}$.} At fixed charge current $I_c$, the NM channel between LBMs encodes the magnitude of the coupling strength between magnets. Extending the channel length $L_{ch}$ reduces the correlation between the magnets due to more spins being neutralized. The polarity of the $J_{ij}$ is controlled through the direction of the input current $I_c$. The analytical analysis is provided in Appendix \ref{Mapping_Charge_to_Spin}, we assume the channel to be Cu with $400$ nm spin-diffusion length.
}
    \label{figure_4}
\end{figure}

The frustrated system in Fig. \ref{figure_3}(d) can be also described by Eq. \eqref{eq: 3LBM-correlation}. All magnets have identical charge inputs (swept from negative to positive values of $I_c$ where positive electron current is measured from the magnet into the channel) leading to a uniform $J_{ij} = \pm J_0$, ferromagnetic/antiferromagnetic interactions. At all currents we observe identical correlations, $i.e.$, $\langle \theta_{12} \rangle = \langle \theta_{13} \rangle = \langle \theta_{23} \rangle$, and thus we only report $\langle \cos\theta_{12} \rangle$ without loss of generality in Fig.~\ref{figure_3}. Similar to the 2-magnet system, positive currents lead to ferromagnetic correlation, however, negative currents result in a saturated correlation of $\cos\theta_{12}=-0.5$ where each magnet (on average) has a 120-degree separation between its neighbors (Fig.~\ref{figure_3}(f)), reminiscent of  frustrated magnets realizing XY models \cite{Berloff_XY_realization}.

\section{Power estimation}
Since the proposed hardware comprises both the neuron and the synapse, the power consumption is expected to be order of magnitude less than possible digital implementations. There are no detailed estimations of the digital footprint for continuous stochastic neurons, however, transistor-level projections indicate that tens of thousands of transistors operating at 10's of $\mu$W's are needed even for \emph{binary} stochastic neurons \cite{kobayashi2023cmos+,debashis2022gaussian}.
To generate significant correlations, our systems presented in Fig.~\ref{figure_3} consumes around $180$ nW and $270$ nW for the 2-LBM and the 3-LBM systems, about 100 nW per LBM terminal. We measure this power as the average at the two extremes of input charge currents for the ferromagnetic and the anti-ferromagnetic cases. These numbers can be understood by relating the necessary spin-currents to charge currents. Spin-currents need to be around $\approx \pm 15 \ I_0 \approx 1.88 \ \mu$A to create large negative or positive correlations. With interface polarizations of $P\approx 0.1$, interface and side channel resistances of $R_{int}$~$\approx$1 \, $\Omega$, $R_{side}$$\approx 0.5$$\Omega$, a spin-diffusion length of 400 nm along 200 nm channels, and resistive division factors diverting the spin-currents, charge currents of around $I_c=250 \ \mu$A are needed (see Appendix \ref{Mapping_Charge_to_Spin}). These charge currents lead to $I_c^2 R$ losses of around 100 nW per LBM arm. Considering the 10-20 $\mu$W estimations for \emph{binary} stochastic neurons for the simpler \emph{Ising} model, the proposed emulator should be at least 3 to 4 orders of magnitude more energy-efficient over digital implementations of the classical Heisenberg model. 
\begin{figure}[t!]
   \centering
    \includegraphics[width=\linewidth]{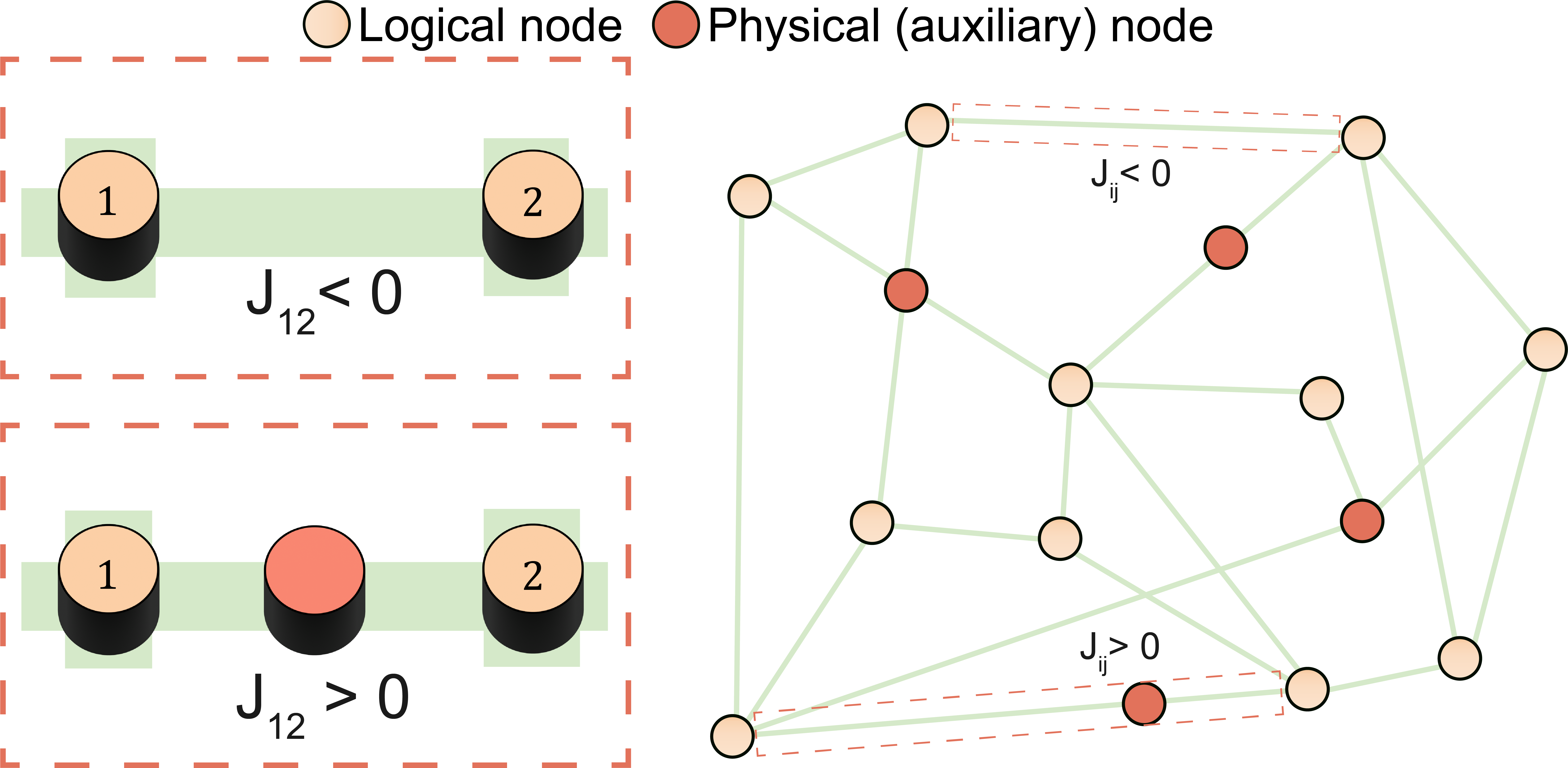}
    \caption{\textbf{Programming $\pm$ weights.} Auxiliary magnets are introduced to choose the polarity of the interaction strength $J_{ij}$, where the input current is fixed such that all magnets have negative correlation ($J_{ij} < 0$). The magnitude of the $J_{ij}$ is tuned by the channel length between any two spins ($\hat m_i, \hat m_j$).}
    \label{figure_5}
\end{figure}

\section{Programmability}
\label{Programmability}

Another practical aspect is the tunability of the interaction strengths, $J_{ij}$. In the proposed hardware, these interactions are described by the amount of spin current received from the other LBMs, which can be tuned through many parameters (e.g., input current, channel material, etc.). In Fig.~\ref{figure_4}, we analyze the programmability of $J_{ij}$ for a channel by tuning the channel length $L_{ch}$, the spin-diffusion length for Cu is around $400$ nm at room temperature \cite{Otani_parameters}, results show a reasonable tunability range for $J_{ij}$. Recently, it was reported that graphene can have a spin-diffusion length of up to 26 $\mu m$ at room temperature \cite{bisswanger2022cvd}. Having longer spin-diffusion length relaxes the constraints on tuning the interaction strength $J_{ij}$ and we can achieve higher values of $J_{ij}$, both are critical for machine learning applications, such as Boltzmann Machines \cite{Hinton_Book_1984,Hinton_assymtric_strength}. Throughout, we assumed configurations where all charge currents have the same sign and magnitude, leading to $J_{ij}$ with the same sign. 

However, systems with frustration involves positive and negative $J_{ij}$. One way to imlement $\pm J_{ij}$ in one configuration is by using auxiliary magnets. We set the network to have all negative weights, and through graph embedding via auxiliary nodes (Fig.~\ref{figure_5}), a mix of positive and negative weights can be obtained, where the role of auxiliary magnets is to have a double negative effect such that $-(-J_{ij})$ = $+J_{ij}$. Keeping in mind that to ensure symmetric $J_{ij}$ values, we inject the same charge current over all LBMs.

From a practical point of view, our programming approach is readily applicable for inference (or sampling) problems, where $J_{ij}$ are fixed and do not need to change. This would also include optimization approaches such as parallel tempering. For optimization problems, where the system needs to be annealed by varying the temperature, e.g., simulated annealing, we can tune the system temperature for a given set of $J_{ij}$ by gradually increasing the injected charge current $I_c$ from a low to high value. For machine learning applications where the system needs to update $J_{ij}$ at each training step, using the same programmability approach would be challenging. One possibility  for training the system without rebuilding the circuit at each training step is to introduce multiple ferromagnetic contacts that can inject currents with varying distances between electrodes.

\begin{figure}[b!]
   \centering
    \includegraphics[width=\linewidth]{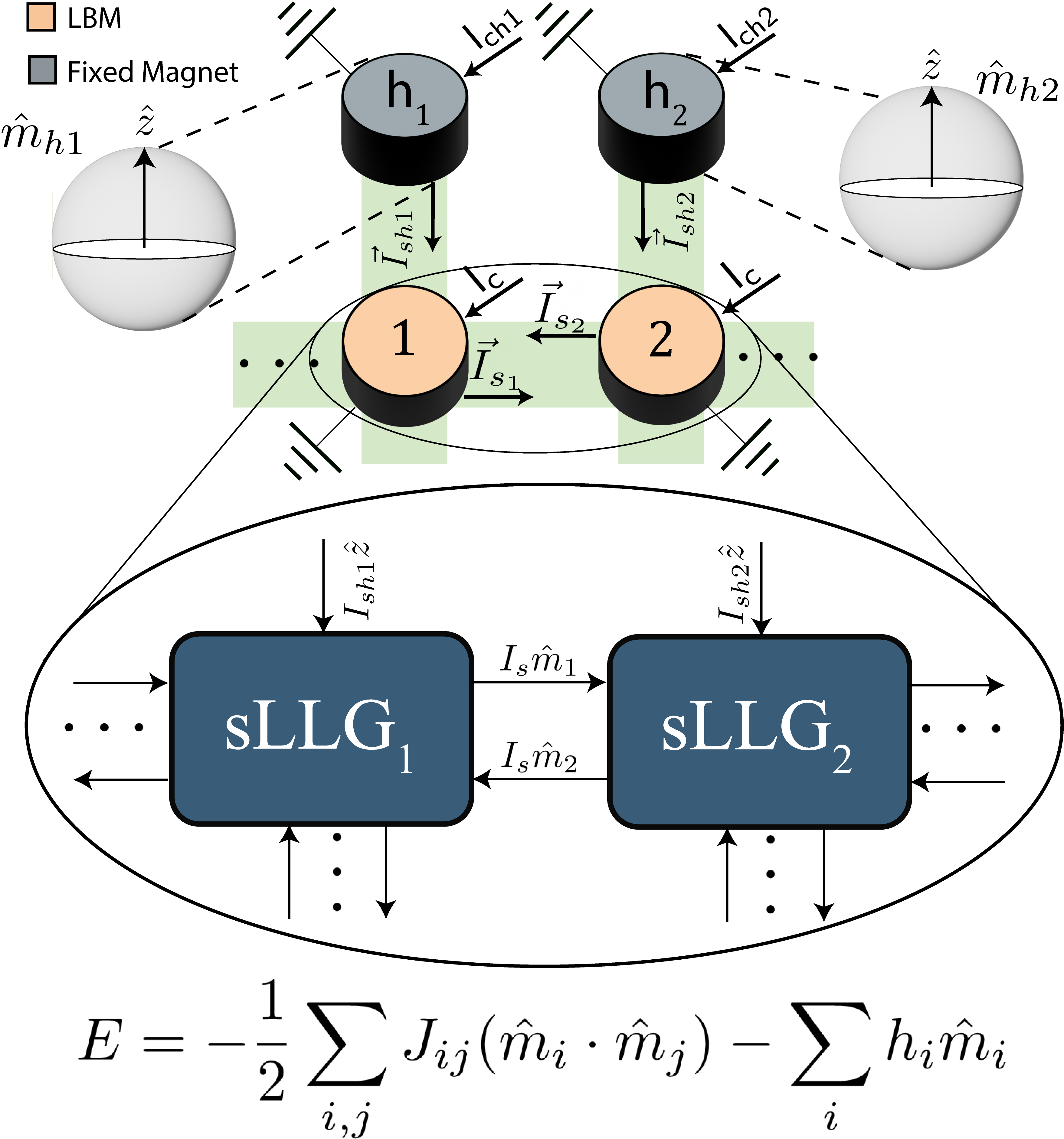}
    \caption{\textbf{Heisenberg machine with bias terms.} For an LBM array, bias terms $h_i$ can be introduced by stable magnets with a fixed magnetization direction $\hat{m}_{hi}$ producing fixed spin current $I_{shi}$ to incident magnets.}
    \label{figure_6}
\end{figure}

\section{Readout of LBMs}
\label{Readout}
An important practical consideration is the ability to read out the correlations induced in LBMs. Advances in modern spintronic capabilities offer various possibilities. We briefly consider three approaches that could be used. First, an experimentally demonstrated approach is using nearby ferromagnets for potentiometric read out, e.g., as commonly used for topological insulators (TI) \cite{hong2012modeling,kim2019electrical}. Secondly, inverse spin Hall effects in heavy metals and TIs can be used via terminals placed near the LBMs \cite{choi2022all,pham2020spin}.  Finally, magnetic tunnel junctions built on top of the LBMs could be used for read-out as commonly used in spin-orbit torque-based structures \cite{liu2012spin}. This approach may require isolated and separate terminals for the MTJ-read out and the charge injection over the LBM similar to those in all-spin logic devices and potentially is more challenging from a fabrication point of view \cite{behin2010proposal}. 

In the schemes we proposed, readout of the full magnetization vector is not possible and the methods we suggest can only read the projection of the magnet state along a specified direction. Even with this loss of information, the continuous nature of spins are preserved and the internal dynamical evolution of the system rely fully on the 3D magnetization vectors.

\begin{figure}[t!]
   \centering
    \includegraphics[width=\linewidth]{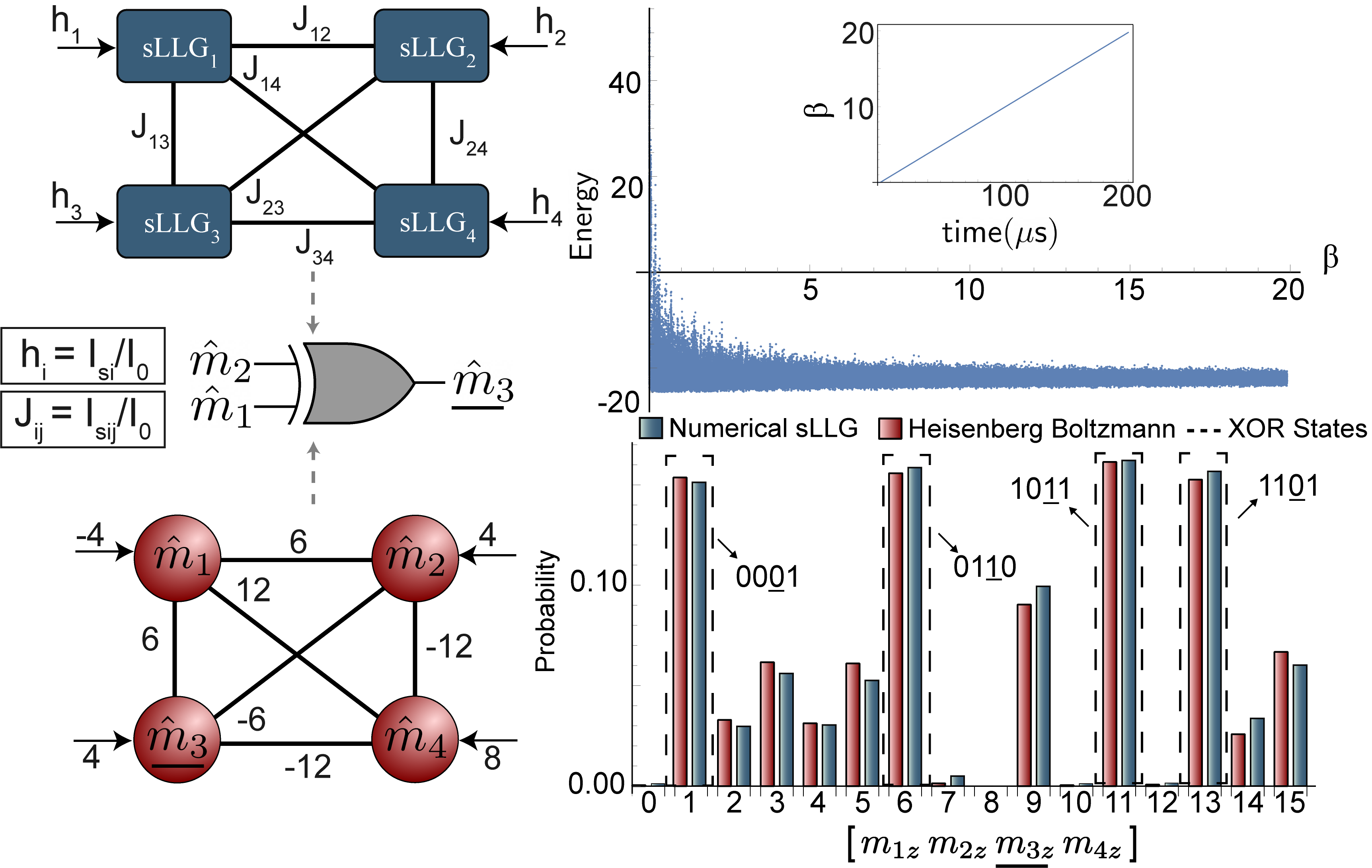}
    \caption{\textbf{Heisenberg XOR Gate.} Training four coupled LBMs modeled by sLLGs ($\hat{m}_4$ is an auxiliary spin). The interaction strength $J_{ij} = I_{sij}/I_0$ and bias terms $h_i = I_{si}/I_0$ (shown in the figure) are learned to implement a Heisenberg XOR gate, continuous spins were binarized by thresholding at zero. The network was annealed with a linear profile of the dimensionless inverse algorithmic temperature $\beta$. $\beta$ linearly scales the weights and the biases in the network. In physical implementations, $\beta$ can be adjusted by changing the injected charge currents through each LBM. Results show saturation around the ground state of the network. The truth table of the XOR was verified numerically by the network response, with the correct states being $\{1,6,11,13\}$, matching the expected response obtained by the Boltzmann law. For the histogram producing the truth table, the network was simulated at a constant $\beta = 4$. $I_0 \approx 125$ nA is fixed for all sLLG equations. }
    \label{figure_7}
\end{figure}

\section{Training Heisenberg Machines}
\label{Training Heisenberg Machines}
The energy model shown in Eq. \eqref{Heisen} did not include bias terms for simplicity. To our system of LBMs, bias terms $h_i$ can be introduced using stable magnets with a fixed magnetization as shown in Fig. \ref{figure_6}. More details about Eq. \eqref{Heisen_bias}, and how it breaks the zero net magnetization that was caused by the symmetry of Eq. \eqref{Heisen}, are provided in Appendix \ref{sec: Bias}.

\begin{equation}
   E = - \frac{1}{2} \sum_{i,j} J_{ij}\,(\hat{m}_i\cdot\hat{m}_j)-\sum_{i}h_i\hat{m}_i
   \label{Heisen_bias}
\end{equation}
These bias terms make the energy model completely general, allowing us to construct networks where arbitrary correlations between the LBMs can be designed. To show this, we train a network of LBMs using a continuous generalization of the contrastive divergence algorithm \cite{hinton2002training,larochelle2007empirical} (see Appendix \ref{sec: Heisenberg_Learning} for details), where a Heisenberg XOR gate based on four coupled LBMs is obtained with three XOR spins and an auxiliary spin (Fig.~\ref{figure_7}). For this example, we used coupled sLLG equations without our full transport models for simplicity. Further, by linearly scaling the weights and biases with the dimensionless inverse algorithmic temperature $\beta$, we perform an annealing of the system energy. In practice, this could be done, for example, by increasing the charge current injected into the LBMs. Note that $\beta$ describes only a universal scaling factor of the weights and biases, it is detached from the actual temperature $T$ that determines the physical parameters $I_0$. Our annealing result aligns well with the ground state, the error in the network minimum state is around $0.01 \%$ after 200 $\mu$s. We also obtain the full truth table for the XOR gate by sampling the network output at fixed $\beta$ for 500 $\mu$s. To generate Boolean output values, continuous spins were binarized using thresholding at the zero point. No thresholding was performed during the simulation, however, all spins were binarized post-simulation. Figure \ref{figure_7} shows that the network response is identical to that one obtained by the Boltzmann law, and in agreement with a probabilistic XOR gate operation. Further details are provided in Appendix \ref{sec: Heisenberg_Learning}. In this example, we showed how the 3D dynamics of the Heisenberg model can be adjusted to build a probabilistic XOR gate with \emph{Boolean} output states. The full continuous nature of our spins could be useful in the areas of stochastic computing where arithmetic operations such as multiplication, division and factorization may be simplified by continuous stochastic neurons \cite{alaghi2013survey}. 

\section{Conclusion}
\label{Conclusion}
This work presents a programmable hardware platform to emulate the classical Heisenberg model using non-local spin valves. With analytical and numerical results, we  establish that the physics of low barrier nanomagnets with perpendicular magnetic anisotropy leads to a coupled system whose steady-state behavior is described by a Boltzmann factor ($\propto \exp(-\beta H))$ where $H$ is the three-dimensional classical Heisenberg Hamiltonian. With the tremendous current interest in building programmable  computing hardware for the \emph{Ising} model, the compact and energy-efficient realization of the classical Heisenberg model whose classical emulation would be much more costly than equivalent Ising systems could be useful for a number of applications. These include training modern Hopfield networks to solving continuous optimization problems. We leave natural extensions of the concept to in-plane magnetic anisotropy (IMA) magnets realizing the classical XY model for future investigation. We envision that spin-circuit networks with LBMs can be extended beyond conservative systems described by an energy, such as Bayesian (Belief) networks with asymmetric network connections \cite{Hinton_assymtric_strength,Bayesian_Damien,Bayesian_appenzeller_2020}. 

 The spin-circuit codes used in this study are  available on GitHub \cite{GithubCode}. \\ 

\section*{Acknowledgment}
We acknowledge support from ONR-MURI grant N000142312708, OptNet: Optimization with p-Bit Networks. The authors are grateful to Shun Kanai, Saroj Dash, Punyashloka Debashis and Zhihong Chen for fruitful discussions.

\appendix

\section{SPIN-CIRCUIT MODULES}
\label{Spin-circuit modules}

\begin{figure*}[t!]
   \centering
    \includegraphics[width=\linewidth]{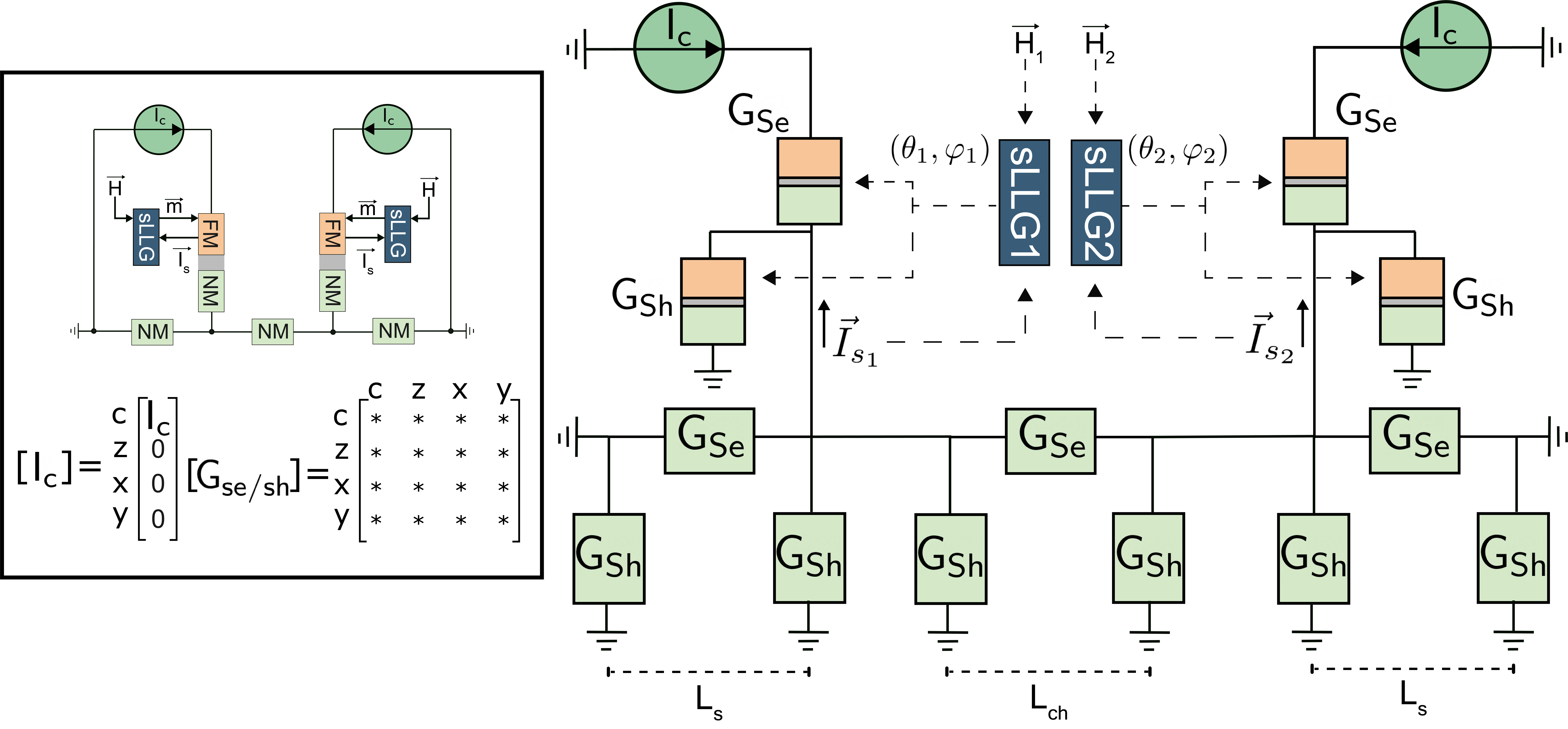}
    \caption{Spin-circuit modules. The LBMs and the spin-neutral channel are described by a combination of series $G_{se}$ and shunt $G_{sh}$ conductance, where each conductance is a $4\times 4$ matrix that takes in consideration the interaction between charges and spins in $z,x,y $ directions. The sLLG module solves the stochastic LLG equation as a function of the received spin current $\vec I_s$ and thermal noise.}
    \label{Figure_8}
\end{figure*}

Spin-circuits \cite{Bauer,srinivasan2013modeling,manipatruni2012modeling,Camsari_modular_approch} provide a generalization of ordinary charge circuits where each node in the circuit is represented by a  4-dimensional voltage, ($[V_c V_z V_x V_y]^T$) corresponding to 3-spin components $(z,x,y)$ and 1-charge component. Figure~\ref{Figure_8} shows the details of the spin-circuit for the 2-LBM system considered in the main paper. For the normal metal  (NM) module, the series conductance $G_{se}$ and the shunt conductance $G_{sh}$ are defined as:
\begin{equation}
G_{se}= 
\left[ \begin{array}{c|cccc}
  & c & z & x & y \\ \hline
c & G_c & 0 & 0 & 0 \\
z & 0 & G_s & 0 & 0 \\
x & 0 & 0&G_s & 0 \\
y & 0 & 0 & 0 & G_s
\end{array} \right]
, % This adds space between the two matrices
G_{sh}= 
\left[ \begin{array}{c|cccc}
  & c & z & x & y \\ \hline
c & 0 & 0 & 0 & 0 \\
z & 0 & G_s' & 0 & 0 \\
x & 0 & 0&G_s'&0 \\
y & 0 & 0 & 0 & G_s'
\end{array} \right]\nonumber
\end{equation}
\noindent where we define $G_c = A_{NM}/(\rho_{NM}L)$, $G_s = A_{NM}/(\rho_{NM}\lambda_s)\csch(L/\lambda_s)$, and $G_s'= A_{NM}/(\rho_{NM}\lambda_s)\tanh(L/2\lambda_s)$. $A_{NM}$ denotes the NM's area, $\rho_{NM}$ is the NM's resistivity, $L$ is the NM' length, and $\lambda_s$ is the NM's spin-diffusion length.  Similarly, the shunt and series conductance for the FM$|$NM interface, if the magnet is pointing in the $+\hat z$ direction, are defined as:

\begin{equation}
G_{se}=
\left[ \begin{array}{c|cccc}
  & c & z & x & y \\ \hline
c & G & P\,G & 0 & 0 \\
z & P\,G & G & 0 & 0 \\
x & 0 & 0&0 & 0 \\
y & 0 & 0 & 0 & 0
\end{array} \right], 
G_{sh}= 
\left[ \begin{array}{c|cccc}
  & c & z & x & y \\ \hline
c & 0 & 0 & 0 & 0 \\
z & 0 & 0 & 0 & 0 \\
x & 0 & 0&a\,G & b\,G \\
y & 0 & 0 & -b\,G & a\,G
\end{array} \right]\nonumber
\label{eq:intz}
\end{equation}
where $G$ is the interface conductance, $P$ is the interface polarization, $a,b$ are the real and imaginary coefficients of the spin-mixing conductance, respectively. The interface conductance can be rotated via $G_{\{sh,se\}} = [U_R]^T  \left[ G_{\{sh,se\}} \right] [U_R]$, where the rotation matrix $[U_R]$ is given by:
\begin{equation}
\resizebox{\columnwidth}{!}{%
$
\left[ \begin{array}{c|cccc}
& c & z & x & y \\ \hline
c& 1&0&0&0\\
z&0& \cos \theta & \sin \theta \cos \phi & \sin \theta \sin \phi \\
x&0& -\sin \theta \cos \phi & \cos \theta + \sin^2 \phi (1 - \cos \theta) & -\sin \phi \cos \phi (1 - \cos \theta) \\
y&0& -\sin \theta \sin \phi & \sin \phi \cos \phi (1 - \cos \theta) & \cos \theta + \cos^2 \phi (1 - \cos \theta)
\end{array} \right]
$%
}\nonumber 
\end{equation}

In our numerical modules, the sLLG module provides the instantaneous magnetization directions for all LBMs $\hat m_i(\theta_i,\phi_i)$, and the circuit simulator rotates the conductances based on the new magnetization. On the other hand, the magnetizations are updated by solving the stochastic LLG based on the received spin currents $\vec I_s$ along with the thermal noise which enters the effective fields $\vec H$.

\begin{equation}
\begin{split}
(1+\alpha^2)\frac{d\hat{m}_1}{dt} = & -|\gamma| \hat{m}_1 \times \vec{H} - \alpha|\gamma| \hat{m}_1 \times (\hat{m}_1 \times \vec{H}) \\
& + \frac{\hat{m}_1 \times (\vec{I_s} \times \hat{m}_1)}{qN_s} + \frac{\alpha (\hat{m}_1 \times \vec{I_s})}{qN_s}
\end{split}
\label{eq: sLLG}
\end{equation}
For LBMs with no effective anisotropy, the magnetic field $\vec H$ is fully characterized by the thermal noise defined as the 3D uncorrelated magnetic fluctuation $\vec H_n$ in the $({z,x, y})$ directions.

\begin{equation}
\text{Var} (H_n^{z, x, y}) = \frac{2\alpha kT}{|\gamma| M_s \mathrm{Vol.} }, \quad  
\mathbb{E} [H_n^{z,x,y}] = 0 
\label{eq: noise}
\end{equation}
For these magnets, at the absence of the thermal noise, the sLLG equation (Eq. \eqref{eq: sLLG}) will be reduced to Eq. \eqref{eq: LLG}. 

The self-consistency between magnetism and transport is well-defined because electronic timescales are much faster than magnetization dynamics, therefore, at each discrete time point, a lumped circuit module for the transport can be defined based on the new magnetizations.

\vspace{-20pt}
\section{MAPPING CHARGE TO SPIN}
\label{Mapping_Charge_to_Spin}

Finding the exact analytical expression for the spin-to-charge mapping is challenging, but in the case of a 2-magnet system, this can be done analytically by a clever coordinate transformation and heavy algebraic manipulation.  We first define a reference frame ($z',x',y'$), where the $+\hat z'$ axis always coincides with the fluctuating direction of the magnet $\hat m_1$. We use this new coordinate system to transform the $(z,x,y)$ coordinates of the channels and the interface matrix of the second magnet. Since the channels we consider in this paper are isotropic in spin (in the absence of any spin-orbit torques or directional spin relaxation), the channel conductances are unaffected by this transformation. For the second magnet, we have $\hat{m}'_2 = R\,\hat{m}_2$: 

\setlength{\arraycolsep}{5pt}
\begin{widetext}
\begin{equation}
R = 
\left[\begin{array}{c|ccc}
& z & x &  y \\ \hline
z & \cos(\theta_1) & \sin(\theta_1)\cos(\phi_1) & \sin(\theta_1)\sin(\phi_1)  \\
x & -\cos(\varphi)\sin(\theta_1) &  \cos(\varphi)\cos(\theta_1)\cos(\phi_1)-\sin(\varphi)\sin(\phi_1) & \cos(\varphi)\cos(\theta_1)\sin(\phi_1)+\sin(\varphi)\cos(\phi_1)  \\
y & \sin(\varphi)\sin(\theta_1) & -\cos(\varphi)\sin(\phi_1)-\sin(\varphi)\cos(\theta_1)\cos(\phi_1) & \cos(\varphi)\cos(\phi_1)-\sin(\varphi)\cos(\theta_1)\sin(\phi_1)
\end{array}\right]
\nonumber
\end{equation}
\end{widetext}

 where the angle $\varphi$ describes the applied rotation around the $(x,y)$ plane. Note that $R$ is described in ($z',x',y'$) rather than ($x',y',z'$), since this is how we express channel and interface matrices in spin-circuit modules described in Eq.~\eqref{eq:intz}. The new coordinates simplify the system for the purpose of finding the charge to spin conversion ratio, since now the coupled system reduces to the simpler configuration where one magnet ($\hat m_1$) is fixed to the $+\hat{z}'$ direction and injects a spin-current of the form $\vec{I}_s = I_s \hat{z}'$. 
 
Another key point that needs to be considered is \emph{extracting} the component of incident spin currents along the direction of \emph{transmitting} magnets. As such, we need to decompose incident spin currents (such as $\vec{I}_{s2}$ in Fig.~\ref{Figure_8}) into its constituents. For this purpose, we choose a commonly used non-orthogonal basis for the 2-magnet system \cite{datta2011voltage} $(\hat{m}_1,\hat{m}_2,\hat{m}_1\times\hat{m}_2)$. This allows us to clearly resolve the individual contributions of each magnet to incident spin currents. This new basis can be described by the transformation matrix, $A$, that turns $(z,x,y)$ to $(\hat{m}_1,\hat{m}_2,\hat{m}_1\times\hat{m}_2)$: 

\setlength{\arraycolsep}{10pt}
\begin{widetext}
\begin{equation}
A = 
\left[\begin{array}{c|ccc}
& \hat{m}_1 & \hat{m}_2 & \hat{m}_1 \times \hat{m}_2 \\ \hline
z & \cos(\theta_1) & \cos(\theta_2) & \sin(\theta_1) \sin(\theta_2) \sin(\phi_2-\phi_1) \\
x & \sin(\theta_1) \cos(\phi_1) & \sin(\theta_2) \cos(\phi_2) & \cos(\theta_2) \sin(\theta_1) \sin(\phi_1) - \cos(\theta_1) \sin(\theta_2) \sin(\phi_2) \\
y & \sin(\theta_1) \sin(\phi_1) & \sin(\theta_2) \sin(\phi_2) & \sin(\theta_2) \cos(\theta_1) \cos(\phi_2) - \sin(\theta_1) \cos(\theta_2) \cos(\phi_1)
\end{array}\right]
\end{equation}
\end{widetext}

Then the spin current component of interest is described by:
\begin{equation}
\begin{split}
\vec{I}_{s2} = (R\,A)^{-1}\vec{I_{s2}}(z',x',y') &= I_{sm1} \, \hat{m}'_1 + I_{sm2} \, \hat{m}'_2 \\
&\quad + I_{s\perp} \,(\hat{m}'_1 \times \hat{m}'_2)
\end{split}
\end{equation}

It may seem that in order to find an analytical expression for $\vec{I}_{s2}$, we must know $(\theta_2,\phi_2)$, since the conductance matrices for the second magnet are a function of $(\theta_2,\phi_2)$. However, when we assume $a = 1$, the spin current component along the direction of $\hat{m}_1$, described by $I_{sm1}$, is \emph{independent} of the instantaneous direction of $\hat{m}_2$. This allows us calculate $I_{sm1}$, however, in our new coordinate system (where $\hat{m}_1=\hat z'$). The spin-current $I_{s2}$ also needs to be expressed in $(z',y',x')$.

After these basis transformations and tedious algebra using standard circuit theory, we solve the 2-magnet system and arrive at the analytical expression (keeping only leading order terms for $P$, since typically $P^2 \ll 1$):
\begin{widetext}
\begin{equation}
\frac{I_{sm1}}{I_c} = \frac{P R_{\text{int}} R_{\text{sp}}}{R_{\text{int}}^2 \text{csch}^2\left(\displaystyle\frac{L_s}{\lambda _s}\right) \sinh \left(\displaystyle\frac{L_{\text{ch}}+2 L_s}{\lambda _s}\right)+2 R_{\text{int}} R_{\text{sp}} \text{csch}\left(\displaystyle\frac{L_s}{\lambda _s}\right) \sinh \left(\displaystyle\frac{L_{\text{ch}}+L_s}{\lambda _s}\right)+R_{\text{sp}}^2 \sinh \left(\displaystyle\frac{L_{\text{ch}}}{\lambda _s}\right)}
\label{eq:Charge_Spin_mapping}
\end{equation}
\end{widetext}

\noindent where $R_{\text{sp}}$ is defined as the resistance of a block of NM with length $\lambda_s$, i.e., $R_{sp}=\rho \lambda_s/A_{NM}$ and $R_{\text{int}}$ is the interface resistance, $1/G$.  $L_{ch}$ and $L_{s}$ are the side and middle channel lengths of the normal metals (NM). 

The total mapping factor for the proposed hardware then becomes $I_{M}=I_0 \,(I_c/I_{sm1})$. We used $I_{M}$ in Fig. \ref{figure_3} to relate our dimensionless Boltzmann models to our full numerical results and have obtained agreement, we also used $I_{M}$ in Fig. \ref{figure_4} where the interaction strength $J_{12} = I_c/I_M$ was analyzed by sweeping $L_{ch}$.

It is instructive to consider limits of Eq.~\eqref{eq:Charge_Spin_mapping}, assuming no spin-relaxation, $\lambda_s \rightarrow \infty$): 
\begin{equation}
\frac{I_{sm1}}{I_c}=P\left(\frac{R_{\text{side}}}{R_{\text{int}}+R_{\text{side}}}\right)\left(\frac{R_{\text{side}}\parallel R_{\text{int}}}{2(R_{\text{side}}\parallel R_{\text{int}})+R_{\text{ch}}}\right)
\label{eq:simple}
\end{equation}
where $R_{\text{side}}$ and $R_{\text{ch}}$ is the charge resistance of the side ($\rho L_s/A_{NM}$) and middle ($\rho L_{ch}/A_{NM}$) channels, respectively. 

In our power calculation for significant saturation, we let $I_c = 15 I_0 \,(I_c/I_{sm1}) =15 I_M$ to obtain the $I_c^2 R$ dissipation, $R = R_{\text{side}}+R_{\text{int}}$, note that the term $(I_c/I_{sm1})$ has a fixed value for this 2-LBM configuration. Pure spin neutral channels can further optimize the power consumption of our proposed device, graphene being a good example \cite{Saroj_Dash_spin_experiment,bisswanger2022cvd}.

Finally, we note that in the above analysis we added a special choice of the mixing conductance ($a=1$) which leads to the $(\theta_2,\phi_2)$ independence of $I_{sm1}$. For arbitrary choices of the mixing conductance, one needs to solve a system of self-consistent equations for every new input $I_c$ such that the $(\theta_2,\phi_2)$ and $\vec{I}_{s2}$ agree.

\begin{figure*}[t!]
   \centering
    \includegraphics[width=\linewidth]{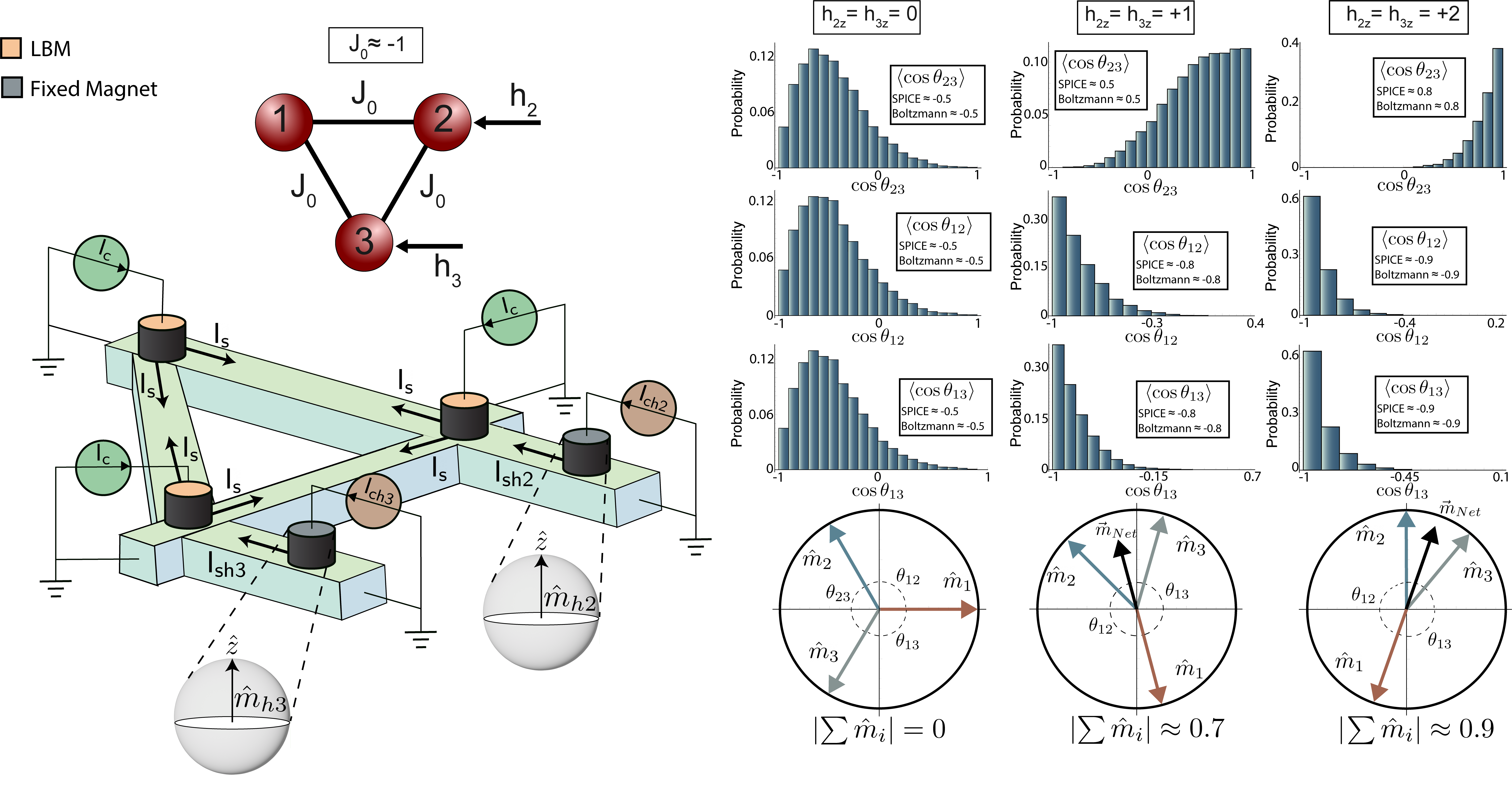}
    \caption{Heisenberg machines with arbitrary correlations. Bias terms can be introduced by using stable magnets to the network. For this example, the interaction strength $J_0$ and and biases $h_i$ are scaled by a factor of $\beta=10$. By projecting the magnetization vectors $m_i$ to a unit circle (approximation not exact), we show that in the absence of biases the LBM array will have zero net magnetization while arrays with biases will have finite net magnetization. General correlations between constituent magnets can be achieved this way.}
    \label{Figure_9}
\end{figure*}

\section{HEISENBERG MACHINE WITH BIAS TERMS}
\label{sec: Bias}
In this section, we discuss how symmetry-breaking bias terms ($h_i$) can be added to the energy model so that we have: 

\begin{equation}
   E = - \frac{1}{2} \sum_{i,j} J_{ij}\,(\hat{m}_i\cdot\hat{m}_j)-\sum_{i}h_i\hat{m}_i
  % \label{Heisen_bias}
\end{equation}

From a hardware point of view, these biases can be implemented using \textit{stable} magnets with fixed magnetization states, rather than LBMs. Accordingly, when a charge current is supplied through one of the fixed magnets, the receiving LBM $\hat{m}_i$ will receive a spin current $I_{shi}$ with a fixed $\hat{m}_{hi}$ direction as we discussed in Fig. \ref{figure_6} in the main text. For subsequent discussions, we assume all fixed magnets are in the $+\hat{z}$ direction though they can in general be arranged to be fixed in an arbitrary direction.

Next we show numerically that with bias terms, an array of LBM can have a finite net magnetization. We examine the bias effect on the frustrated system of three magnets, presented before in the main text (Fig. \ref{figure_3}(d)). Originally, the system had no biases and where the net magnetization was zero due to the  $\pm$ symmetry of the system, as shown in Fig. \ref{Figure_9}. We add bias terms to two of the LBMs at the same inverse temperature where we observe a finite net magnetization (Fig. \ref{Figure_9}). In general, desired and arbitrary correlations between constituent magnets can be achieved by a judicious choice or training of these weights, which we discuss next.

\section{TRAINING HEISENBERG MACHINES: PROOF OF CONCEPT}
\label{sec: Heisenberg_Learning}

In this section, we show how to train Heisenberg machines through an appropriate generalization of the contrastive divergence algorithm, typically used to train Boltzmann machines with binary stochastic neurons. As a representative example, we choose an XOR gate (shown in Fig. \ref{figure_7}). To obtain Boolean states, we take measurements along the $z$-axis, and we binarize continuous spins by setting ${m_i}_z=0$ as the thresholding point between 0 and 1.

The thresholded probabilities of any logical state can then by obtained using the Boltzmann law by integrating over the corresponding range for all spins. For example, for a system of two continuous spins the corresponding probabilities for the four possible states $\{00,01,10,11\}$  are evaluated as follows:\vspace{-15pt}

\begin{widetext}
\begin{equation}
P_{ij}= \int_{\phi_1=0}^{\phi_1=2\pi}\int_{\phi_2=0}^{\phi_2=2\pi}\int_{\theta_1=a}^{\theta_1=b}\int_{\theta_2=c}^{\theta_2=d} {\frac{1}{Z} \exp\left({J_{12} \ \hat{m}_1 \cdot \hat{m}_2+ h_1\hat{m}_1 +h_2\hat{m}_2}\right)\sin(\theta_1)}\sin(\theta_2)d\theta_1d\theta_2d\phi_1d\phi_2
\end{equation}
\end{widetext}

where the states are read as $[{m_1}_z,{m_2}_z]$. In this equation $i,j \in \{0,1\}$ and $(a,b)$ or $(c,d) \rightarrow (0,\pi/2)$ when $i$ or $j$ is 1, and $(a,b)$ or $(c,d) \rightarrow (\pi/2,\pi)$ when $i$ or $j$ is 0. For the partition function $Z$, no special adjustment is required, it is evaluated by doing the normal full integral over the four variables such that the summation of all probabilities add to one. 

The update rules for the weights ($J_{ij}$) and biases ($h_i$) can be obtained by generalizing the  contrastive divergence algorithm commonly used for binary stochastic neurons \cite{hinton2002training,larochelle2007empirical}:

\vspace{-15pt}
\begin{equation}
J_{ij} = \displaystyle J_{ij}+\varepsilon\Bigl(\langle {m}_{iz}{m}_{jz} \rangle_{\text{data}} - \langle {m}_{iz}{m}_{jz} \rangle_{\text{model}}\Bigl)-\varepsilon\,\lambda \, J_{ij}
    \label{eq:Jij_learning}
\end{equation}
\vspace{-10pt}
\begin{equation}
h_{i} = \displaystyle h_{i}+\varepsilon\Bigl(\langle {m}_{iz}\rangle_{\text{data}} - \langle {m}_{iz}\rangle_{\text{model}}\Bigl)-\varepsilon\,\lambda \, h_{i}
    \label{eq:hi_learning}
\end{equation}
where $\varepsilon$ is the learning rate, $\lambda$ is the regularization factor. The correlation of the positive phase $\langle {m}_{iz}{m}_{jz} \rangle_{\text{data}}$ corresponds to clamping the spins to the XOR truth table, while the negative phase correlation $\langle {m}_{iz}{m}_{jz} \rangle_{\text{model}}$ refers to inference stage. Just as in the typical contrastive divergence algorithm, the correlations can be obtained with probabilistic sampling, though in this case we obtained the correlations exactly using the Boltzmann law at each iteration. At the end of training, we performed minor fine tuning of the weights to get near-integer values. In our XOR example, we defined the correct states to be $\{1,6,11,13\}$, such as $(m_{1z},m_{2z},m_{3z})$ are the two inputs and the output of the XOR gate, respectively. The fourth spin $m_{4z}$ is an auxiliary / hidden spin state. 
\vspace{-10pt}

\section{PHYSICAL PARAMETERS}
\label{Physical Parameters}
\vspace{-20pt}
% Table
\begin{table}[h!]
\centering
\scriptsize % Further reduce font size
\begin{tabular}{@{}p{5.5cm}p{2.3cm}p{0.6cm}@{}}
\toprule
\textbf{Parameter}                & \textbf{Value} & \textbf{Unit} \\
\midrule
Interface polarization ($P$)                  & 0.1            & -- \\
Gilbert damping coefficient ($\alpha$) & 0.01           & -- \\
Saturation magnetization ($M_s$)     & 1100 $\times 10^3$           & A/m \\
Magnet volume ($\mathrm{Vol}.$)               & (30 $\times$ 30 $\times$ 2)            & nm$^3$ \\
Interface conductance ($G$)         & 1             & S \\
Spin-Mixing conductance real part ($a\,G$)     & 1              & -- \\
Spin-Mixing Conductance Imaginary Part ($b\,G$)     & 0              & -- \\
\midrule
NM Spin-Diffusion Length ($\lambda_{s}$) & 400             & nm \\
NM Resistivity ($\rho_{NM}$)                    & 2.35            & $\mu\Omega\cdot$cm \\
NM Length ($L = L_s = L_{ch}$)                 & 200            & nm \\
NM Area ($A_{NM}$)  & 1.11 $\times 10^{-14}$             & m$^2$ \\
\midrule
Temperature                       & 300            & K \\
Transient Time Step (SPICE)       & 10            & ps \\
\bottomrule
\end{tabular}
\label{Parameters_sub}
\end{table}
 The physical parameters we used in our simulations are reported in the Table above, where the parameters of the NM channel are chosen according to experiments performed in metallic non-local spin vales \cite{Otani_parameters}. The transient noise (.trannoise) function of HSPICE has been used to solve the stochastic differential equations. This solver has been rigorously benchmarked with exact time-dependent Fokker-Planck equations in Ref.~\cite{torunbalci2018modular}.

\end{document}